\documentclass[12pt]{article}
\pdfoutput=1
\usepackage{amsmath}
\usepackage{amssymb}
\usepackage{graphicx}
\usepackage{subcaption}
\usepackage{url}
\usepackage{comment}

\usepackage[sort&compress, numbers, merge]{natbib}

\usepackage[colorlinks=true, linkcolor=blue, citecolor=blue,
urlcolor=black]{hyperref} 

\title{Indirect Probe of Electroweak-Interacting Particles with Mono-Lepton Signatures at Hadron Colliders
}
                 
\begin{document}
\baselineskip 0.6cm

\def\simgt{\mathrel{\lower2.5pt\vbox{\lineskip=0pt\baselineskip=0pt
           \hbox{$>$}\hbox{$\sim$}}}}
\def\simlt{\mathrel{\lower2.5pt\vbox{\lineskip=0pt\baselineskip=0pt
           \hbox{$<$}\hbox{$\sim$}}}}
\def\simprop{\mathrel{\lower3.0pt\vbox{\lineskip=1.0pt\baselineskip=0pt
             \hbox{$\propto$}\hbox{$\sim$}}}}
\def\tr{\mathop{\rm tr}}
\def\SU{\mathop{\rm SU}}

\begin{titlepage}

\begin{flushright}
IPMU18-0176 \\
\end{flushright}

\vskip 1.1cm

\begin{center}

{\Large \bf 
Indirect Probe of Electroweak-Interacting Particles with Mono-Lepton Signatures at Hadron Colliders
}

\vskip 1.2cm

Shigeki Matsumoto,
Satoshi Shirai and
Michihisa Takeuchi
\vskip 0.5cm

{\it
{Kavli Institute for the Physics and Mathematics of the Universe (WPI), \\The University of Tokyo Institutes for Advanced Study, The University of Tokyo, Kashiwa
 277-8583, Japan}
}

\vskip 1.0cm

\abstract{
Electroweak-interacting massive particles (EWIMPs) exist in a broad class of new physics models beyond the Standard Model. Searching for such particles is one of most primary goal at the LHC and future colliders.
The  EWIMP generally affects the LHC signatures through quantum corrections even without direct productions. By measuring the Standard Model processes precisely, we can indirectly probe the EWIMPs. In this paper, we study the current constraint and future prospect of the EWIMPs by using the precision measurements of mono-lepton production from the charged Drell-Yan processes at hadron colliders. We found the mono-lepton signature can be a better probe than dilepton signature from the neutral Drell-Yan processes.
}

\end{center}
\end{titlepage}

\section{Introduction}
\label{sec:intro}
An electroweak-interacting massive particles (EWIMP) is quite generic prediction of various extensions of the Standard Model (SM).
Supersymmetric (SUSY) models and extra dimension models predict the massive partner particles whose gauge charges are identical to the SM particles.
A generic class of the extended Higgs models also contain the scalar particles which have electroweak charges.

The electroweak gauge interaction also plays a critical role in the dark matter physics.
In many new physics models, the dark matter abundance is controlled by the electroweak gauge interaction.
The most drastic example of the electroweak-interacting dark matter is so-called ``minimal dark matter" (MDM) model \cite{Cirelli:2005uq, *Cirelli:2007xd, *Cirelli:2009uv}.
The idea of this model is that a large electroweak gauge charge can automatically stabilize a particle without any ad-hoc $Z_2$ parity.
In this case, a 5plet Majorana fermion or 7plet real scalar particle is a candidate of the dark matter.

Searching for such particles are essential to reveal the nature of fundamental physics.
One of the most important goal of the LHC is to discover such particles.
In fact, the search strategy of the EWIMPs at the LHC is not straightforward.
The LHC signatures of the EWIMP production strongly depend on how the produced particles decay.
If the EWIMP decays into much lighter particles and the decay products contain leptons, photons or missing energy, the LHC reach is getting excellent.
For the SUSY case, the mass constraint on the chargino and neutralino can be now better than 1 TeV, if the decay products contain multi-leptons \cite{Sirunyan:2017lae, Aaboud:2018zeb}.
The constraint of charginos/neutralinos decaying into photons is also severe \cite{Aaboud:2018doq} and this can exclude a broad class of the low-scale gauge mediation models \cite{Lu:2017oee}.

If the mass difference between the EWIMP and daughter particle is small,
the EWIMP cannot emit energetic particles.
However, due to phase space suppression,  the decay rate of the EWIMP is suppressed and can be long-lived.
In such a case, the EWIMPs provide exotic charged tracks.
A prime example is the Wino dark matter in the SUSY model \cite{Ibe:2006de, *Buckley:2009kv, Asai:2007sw, *Asai:2008sk, *Asai:2008im}.
The Wino dark matter candidate predicts rich signatures in the cosmic rays and direct detection experiments 
\cite{Hisano:2003ec, *Hisano:2004ds, *Hisano:2005ec, Hisano:2010fy, *Hisano:2010ct, *Hisano:2012wm, *Hisano:2015rsa}.
The Wino dark matter is motivated very well from the viewpoint of SUSY model building, which a generic prediction from the anomaly mediation\,\cite{Randall:1998uk, Giudice:1998xp}. This framework is compatible with the so-called ``mini-split SUSY'' scenario\,\cite{Wells:2003tf, *Wells:2004di, ArkaniHamed:2004fb, *Giudice:2004tc, *ArkaniHamed:2004yi, *ArkaniHamed:2005yv}, and the discovery of the 125\,GeV Higgs boson\,\cite{Aad:2012tfa, Chatrchyan:2012ufa} triggers this framework to attract more and more attention\,\cite{Hall:2011jd, *Hall:2012zp, *Nomura:2014asa, Ibe:2011aa, *Ibe:2012hu, Arvanitaki:2012ps, ArkaniHamed:2012gw}.
In such a mini-split SUSY, the decay length of the charged Wino is around 6 cm \cite{Ibe:2012sx, McKay:2017rjs} and the mass constraint of the Wino is around 500 GeV \cite{Aaboud:2017mpt}.
Recently, several ideas to catch a shorter charged track are proposed \cite{Mahbubani:2017gjh,Fukuda:2017jmk}, and Higgsino dark matter in high-scale SUSY \cite{Nagata:2014wma} and MDM, whose charged track is much shorter, can be tested by improving tracking technique and detectors.

The hardest EWIMP target at the LHC is the case that the decay products are soft and there no exotic charged tracks. 
Such cases have been intensively studied and various LHC signatures to increase the sensitivity of the EWIMP search are proposed
\cite{Gori:2013ala, Han:2013usa, Han:2014kaa, Bramante:2014dza, Baer:2014kya, Bramante:2014tba, Ismail:2016zby}.
The aim of this work is propose another strategy for the EWIMP search.

As far as a particle has electroweak gauge charge, it inevitably affects the ordinary SM process through the quantum corrections.
With the precision measurement of the SM process enable us to probe the EWIMP indirectly, even without direct EWIMP production.
In the previous works \cite{Harigaya:2015yaa, Matsumoto:2017vfu}, we propose indirect probes of the EWIMPs, by using the precision measurement of the dilepton distribution at  lepton colliders and  hadron colliders.
Along this line, the prospect at a future 100 TeV hadron collider has been studied \cite{Chigusa:2018vxz}.
In this paper, we extend the previous analysis to mono-lepton signatures.
We find that the mono-lepton signatures at the hadron collider is also useful probe of the EWIMPs and potentially its sensitivity is better than  the dilepton signatures.

\section{Correction from EWIMPs to Mono-lepton Signatures}
In the Standard Model, the mono-lepton and missing energy signatures come from a virtual $W$ boson exchange.
This process has a reasonably large cross section, and is often utilized for a heavy $W'$ search.
Instead of direct production of new physics particles, we focus on the the quantum effect from the EWIMPs.
The EWIMP loop contribution affects the $W$ boson propagator and accordingly the mono-lepton signatures are modified.

\begin{figure}[h]
\centering
\includegraphics[clip, width = 0.5 \textwidth]{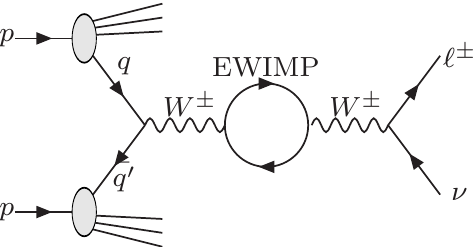} 
\caption{Radiative correction from EWIMPs.}
\label{fig:FG}
\end{figure}

Here we discuss the radiative corrections to the mono-lepton signatures form the EWIMPs.
Fig.~\ref{fig:FG} shows an example of the one-loop correction to the Drell-Yan processes.
The interference between the EWIMP loop and the Standard Model processes  affects the cross section of the mono-lepton productions.
After integrating the EWIMP field out from the original Lagrangian at one-loop level, we obtain the following effective Lagrangian for the mono-lepton signatures:
\begin{align}
	\mathcal{L}_\mathrm{eff} = \mathcal{L}_\mathrm{SM}
	+  \frac{d_{WW}}{4} W^a_{\mu\nu}\,\Pi(-D^2/m^2)\,W^{a \mu \nu}
	+ \cdots\,,
	\label{eq: effective lagrangian}
\end{align}
where $\mathcal{L}_\mathrm{SM}$ is the SM Lagrangian, $m$ is the EWIMP mass, $g$  is the gauge coupling of SU(2)$_{\rm L}$ and $W^a_{\mu\nu}$  the field strength tensor, respectively, with $D$ being the covariant derivative acting on $W^{a \mu \nu}$.
The $\Pi$ is a loop function of renormalized self-energy of the $W$ boson and given by
\begin{align}
\Pi(x)  = 
\begin{cases}
\dfrac{1}{16\pi^2} \displaystyle\int_0^1 dy\,y (1 - y) \ln [1 - y(1-y)\,x]  & (\mbox{fermion})
\\
\dfrac{1}{16\pi^2} \displaystyle\int_0^1 dy\,(1 - 2y)^2 \ln [1 - y(1-y)\,x] & (\mbox{scalar})
\end{cases}\,.
\end{align}
Here we use the $\overline{\mathrm{MS}}$ regularization scheme with the renormalization scale $\mu = m$.
The gauge factor $d_{WW}$ is given by
\begin{align}
d_{WW} = \frac{g^2 n(n-1)(n+1)}{12}
\begin{cases}
1 & ({\rm complex~scalar})\\
8 & ({\rm Dirac~fermion})  
\end{cases}.
\label{eq:gaugefactor}
\end{align}
In the cases of a Majorana fermion and real scalar, another factor $1/2$ should be multiplied.
If the EWIMP has other internal degrees of freedom such as color factor, this factor should be also multiplied.

The matrix element of the mono-lepton process at leading order (LO) is
\begin{align}
	{\cal M}_{\rm SM} [ u(p) \bar{d}(p') \to \ell^+(k) {\nu}(k') ] = \frac{g^2}{2}
	\frac{[\bar{v}_L( d;p') \gamma^\mu  u_L(u;p)]\,[\bar{u}_L( \ell;k ) \gamma_\mu v_L( \nu;k' )] }
	{\hat{s} - m^2_W},
	\label{eq: LO}
\end{align}
The amplitude of the one-loop diagram with the EWIMP is:
\begin{align}
	{\cal M}_{\rm EWIMP} [ u(p) \bar{d}(p') \to \ell^+(k) {\nu}(k') ] = \frac{g^2}{2}
	\frac{[\bar{v}_L( d;p') \gamma^\mu  u_L(u;p)]
	\,d_{WW}\,\hat{s}\,\Pi(\hat{s}/m^2)\,
    [\bar{u}_L( \ell;k ) \gamma_\mu v_L( \nu;k' )] }
	{(\hat{s} - m^2_W)^2},
	\label{eq: NLO}
\end{align}

With the EWIMP loop, the cross section is modified as:
\begin{align}
\frac{\hat{\sigma}_{\rm EWIMP} - \hat{\sigma}_{\rm SM}}{  \hat{\sigma}_{\rm SM}}
= \frac{ |{\cal M}_{\rm EWIMP}+{\cal M}_{\rm SM}|^2-|{\cal M}_{\rm SM}|^2}{|{\cal M}_{\rm SM}|^2
}
 \simeq 2 d_{WW} \frac{\hat{s} {\rm Re}(\Pi(\hat{s}/m^2))}{\hat{s} -m_{W}^2}. \label{eq:correction}
\end{align}

\begin{figure}[h]
\centering
\includegraphics[clip, width = 0.7 \textwidth]{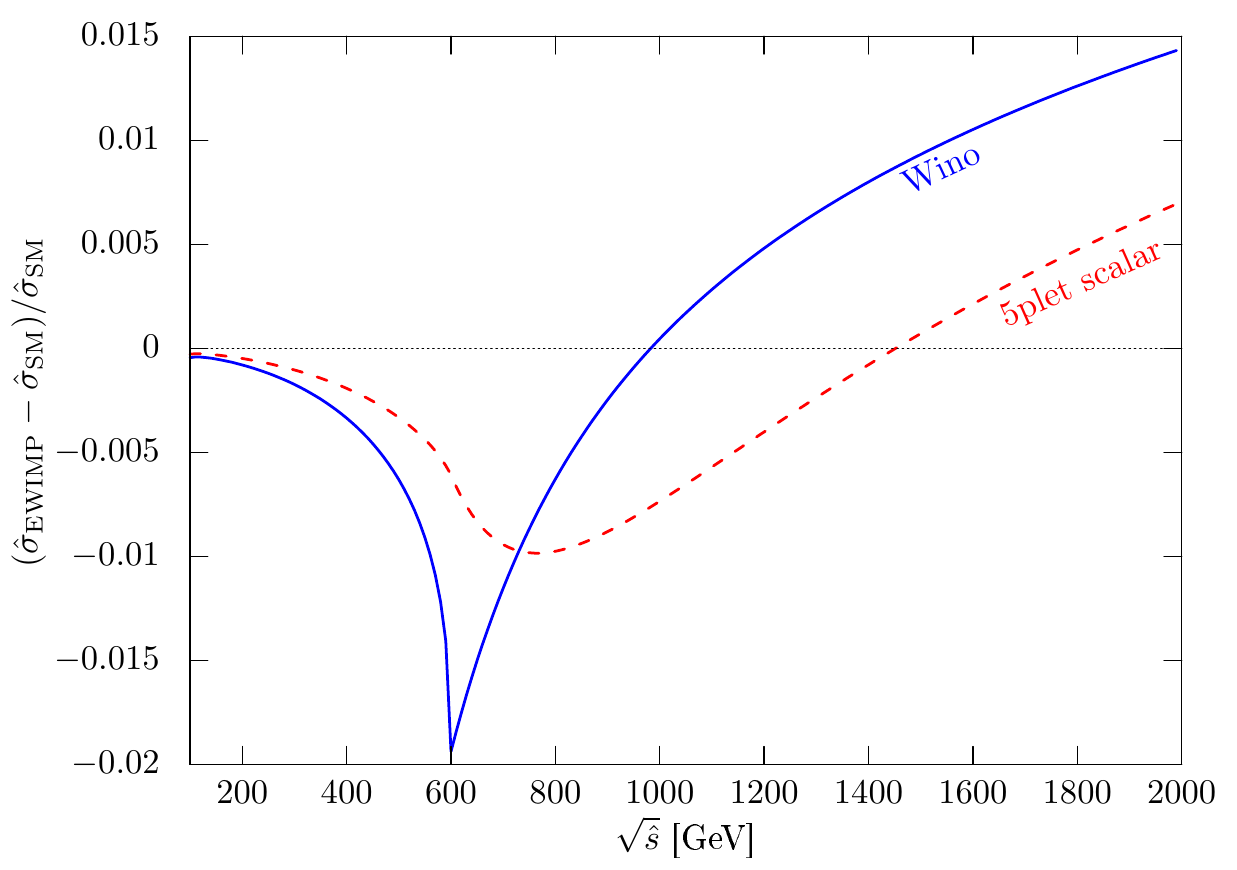} 
\caption{The EWIMP correction to the partonic cross section.
The blue and solid line shows the corrections from the 300 GeV Wino ($n=3$, Majorana fermion) and red and dashed 300 GeV 5plet real scalar.
}
\label{fig:dsigma}
\end{figure}

In Fig.~\ref{fig:dsigma}, we show the modification of the partonic cross section as a function of the center of the center-of-mass energy of the initial state partons $\sqrt{\hat{s}}$ or the invariant mass of the charged lepton and neutrino $m_{\ell \nu}$.
As examples, we show the cases of 300 GeV Wino ($n=3$, Majorana fermion) and 5plet real scalar particles.
For the smaller $m_{\ell \nu} \ll 2 m$, the correction is suppressed and proportional to $m_{\ell \nu}^{2}/m^2$.
When $m_{\ell \nu} \simeq 2 m$, the correction negatively contributes and reduced the cross section by a factor of $d_W/(18\pi^2)$ for the fermion and  $d_W/(36\pi^2)$ for the scalar case.
In the cases of the Wino and 5plet real scalar, these factors are 0.019 and 0.006 respectively, as seen in the figure.
For the larger $m_{\ell \nu} \gg 2 m$, the cross section gets logarithmically enhanced, since the EWIMP particles strengthen the gauge coupling $g$, as the EWIMPs increase the coefficient of the ${\rm SU}(2)_{\rm L}$ gauge beta function.

\begin{figure}[h]
\centering
\includegraphics[clip, width = 0.7 \textwidth]{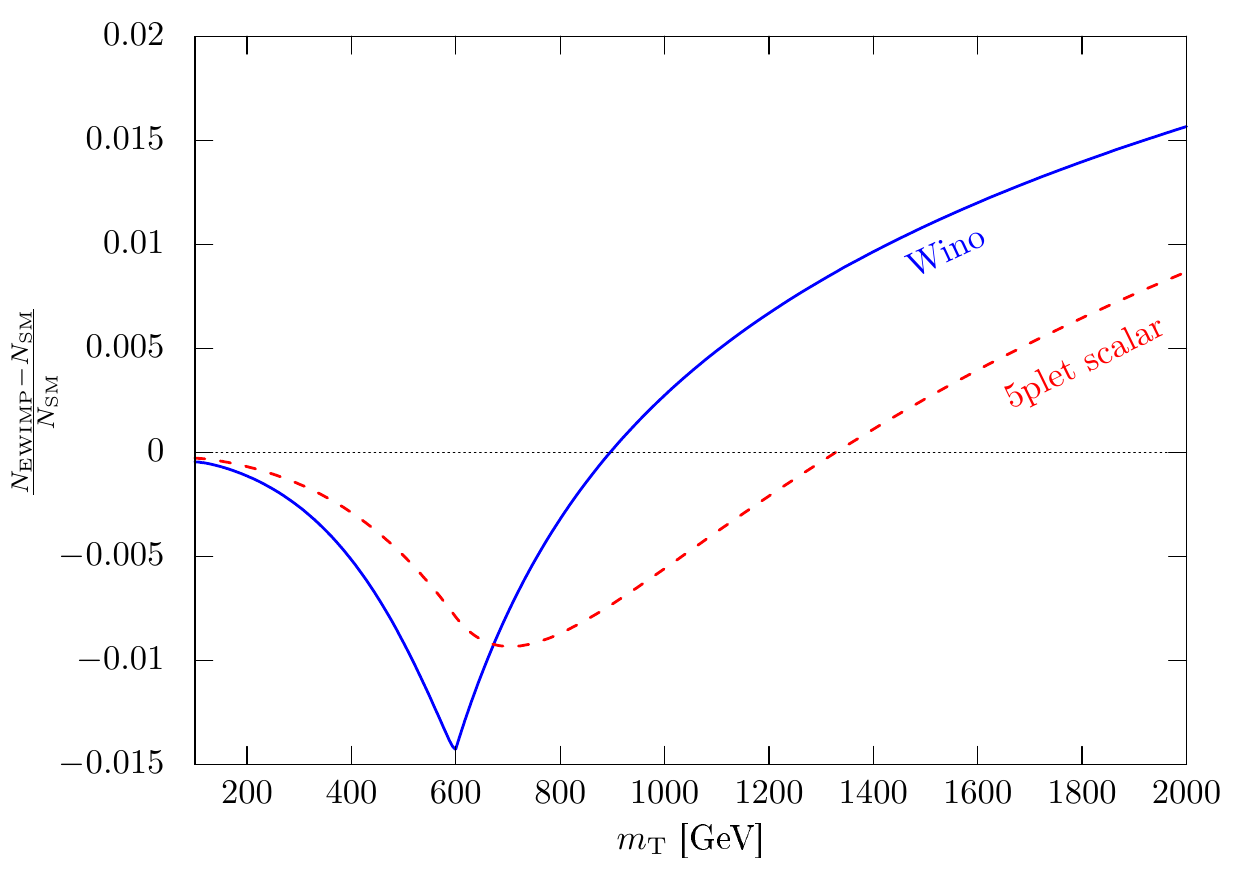} 
\caption{
The EWIMP correction to the transverse mass distribution at the 13 TeV running LHC.
The blue and solid line shows the corrections from the 300 GeV Wino and red and dashed 300 GeV 5plet real scalar.}
\label{fig:dmtsigma}
\end{figure}

In reality, for the mono-lepton signatures, we cannot fully reconstruct the invariant mass of the charged lepton and neutrino, as the longitudinal momentum of the neutrino cannot be measured.  
It is useful to use the transverse mass,
\begin{align}
m^2_{\rm T} = 2 p_{{\rm T}, \ell} p_{\rm T, miss} (1-\cos( \phi_{{\rm T}, \ell,{\rm miss}} )),
\end{align}
where $\phi_{{\rm T}, \ell,{\rm miss}}$ is the azimuthal angle
between the lepton and the missing energy directions in the transverse plane.
The transverse mass $m_{\rm T}$  is less than $m_{\ell \nu}$ and its distribution has a so-called Jacobian peak at $m_{\rm T} = m_{\ell \nu}$.
In Fig.~\ref{fig:dmtsigma}, we show the EWIMP correction to the transverse mass distribution for 13 TeV LHC running.
To estimate this correction, we generate the SM mono-lepton signals, by using the program MG5\_aMC@NLO~\cite{Alwall:2014hca,*Alwall:2011uj}.
We reweight events by following Eq.~\eqref{eq:correction}, using the invariant mass of the parton-level lepton and neutrino.
In Fig.~\ref{fig:dmtsigma}, we show the EWIMP correction to the transverse mass distribution at the 13 TeV LHC.
We see that even in the transverse momentum distribution, the EWIMP corrections can remain and the size of the correction is almost same as the correction to the $m_{\ell \nu}$ distribution. 
In the following, we discuss how we can test this deviation from the Standard Model case.

\section{Analysis of LHC Signatures}
Here we compare the EWIMP signatures with current observation \cite{Aaboud:2017efa} and study the future prospects.
In the following analysis, we perform the Monte-Carlo simulation with programs MG5\_aMC@NLO event generator~\cite{Alwall:2014hca,*Alwall:2011uj}, interfaced to Pythia 6~\cite{Sjostrand:2006za} and Delphes 3~\cite{deFavereau:2013fsa,*Ovyn:2009tx}.

The EWIMP signal, $dN_{\rm EWIMP}/dm_{\rm T}$, mainly come from the interference between the SM charged Drell-Yan process and the EWIMP one-loop diagram. In order to take into account the effect of higher order QCD correction, detector effect, kinematical selection and so on, we calculate $dN_{\rm EWIMP}/dm_{\rm T}$ by multiplying the factor $(\hat{\sigma}_{\rm BSM}- \hat{\sigma}_{\rm SM})/\hat{\sigma}_{\rm SM}$ obtained in the previous section to the $W$ originated background number estimated by the ATLAS collaboration \cite{Aaboud:2017efa} at each bin.

\begin{figure}[t]
	\centering
	\subcaptionbox{\label{fig:e}Electron channel}{\includegraphics[width=0.47\textwidth]{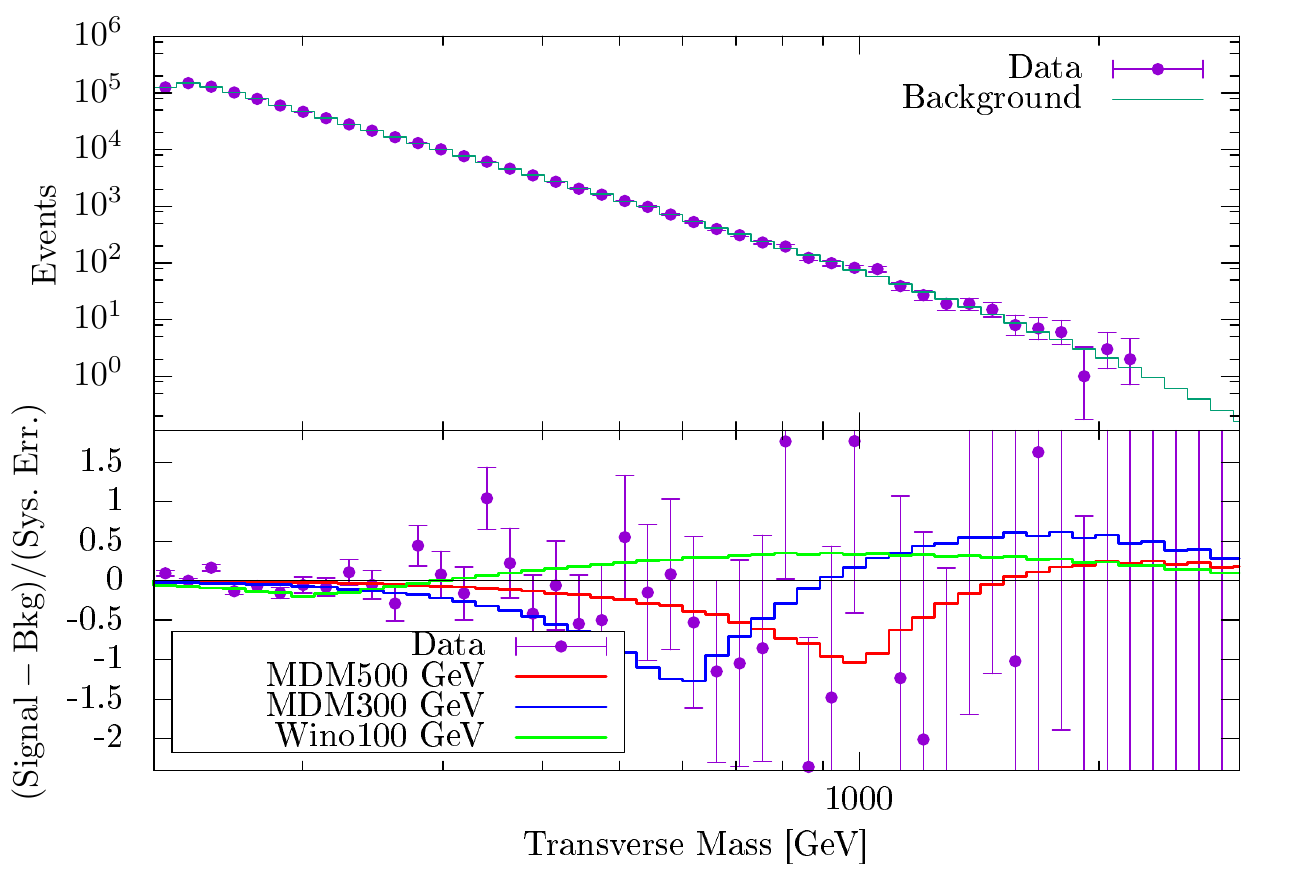}}
	\qquad
	\subcaptionbox{\label{fig:m}Muon channel}{\includegraphics[width=0.47\textwidth]{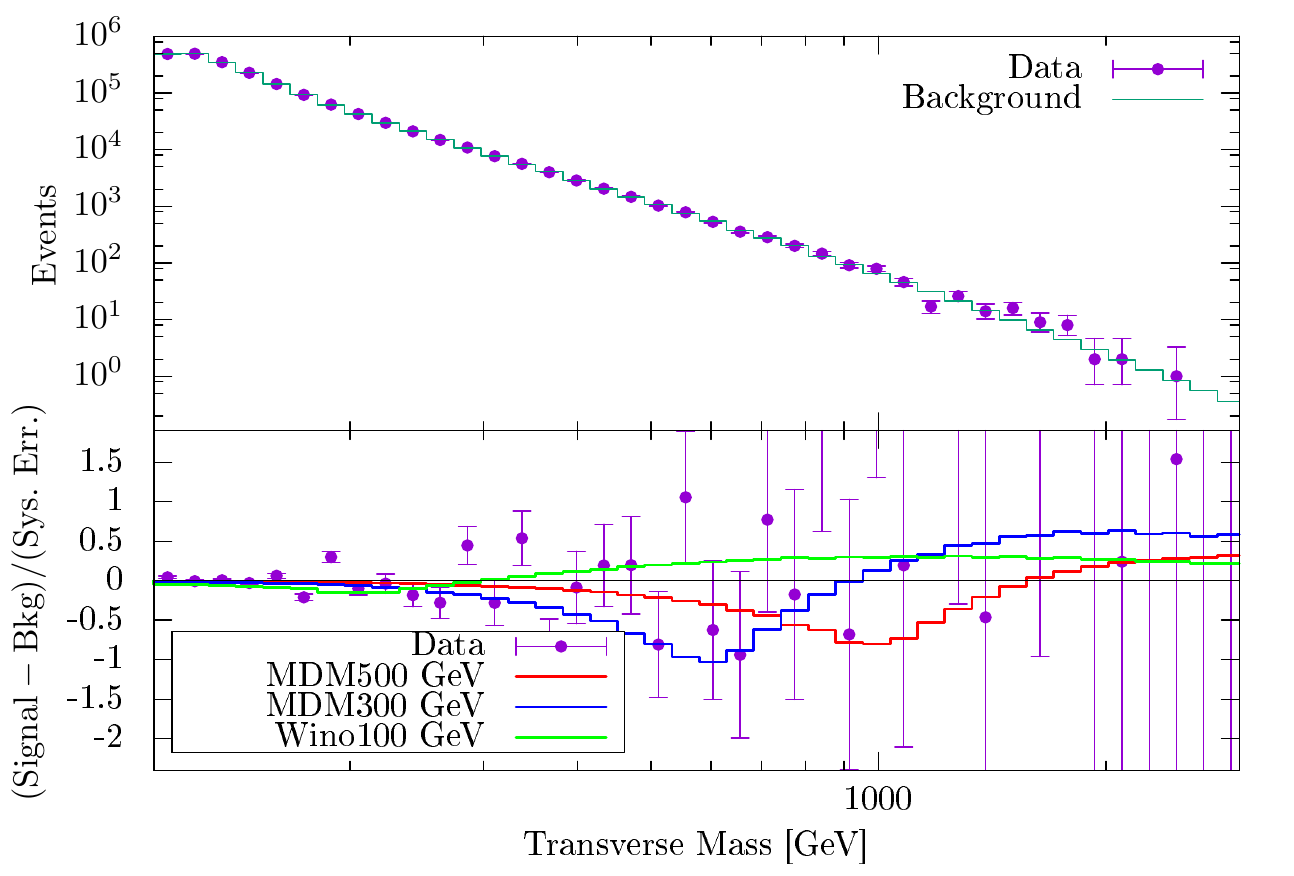}}
\caption{
The observed data and background of the transverse mass distribution at the ATLAS detector with 13 TeV running and integrated luminosity $36~{\rm fb}^{-1}$ \cite{Aaboud:2017efa}.
In the lower panel, we show the pull of the data from the background and the EWIMP signatures of the 100 GeV Wino and 300 GeV and 500 GeV 5plet Majorana fermion (MDM).
}
\label{fig:data}
\end{figure}

In Fig.~\ref{fig:data}, we show the observed data and background of the transverse mass in the mono-lepton events \cite{Aaboud:2017efa}.
We show the difference of the data from the background, divided by the systematic uncertainty (pull) in the lower panel.
We also show the EWIMP signatures of the 100 GeV Wino and 300 GeV and 500 GeV 5plet Majorana fermion (MDM).

So far the observed data and the background are consistent, and we can set the limit on the EWIMP contribution.
For this purpose, we need to know the details of the systematic uncertainties of the Standard Model background prediction.
However, we cannot get detailed information on the systematic uncertainties with public data and we simply assume that systematic uncertainty of each bin is independent.
Under this assumption, we compose the $\chi^2$ variable and set the limit on the mass and charge of the EWIMP (the number of $\SU(2)_{\rm L}$ quantum charge, $n$).

\begin{figure}[t]
	\centering
	\subcaptionbox{\label{fig:result_fermion} Majorana fermion }{\includegraphics[width=0.47\textwidth]{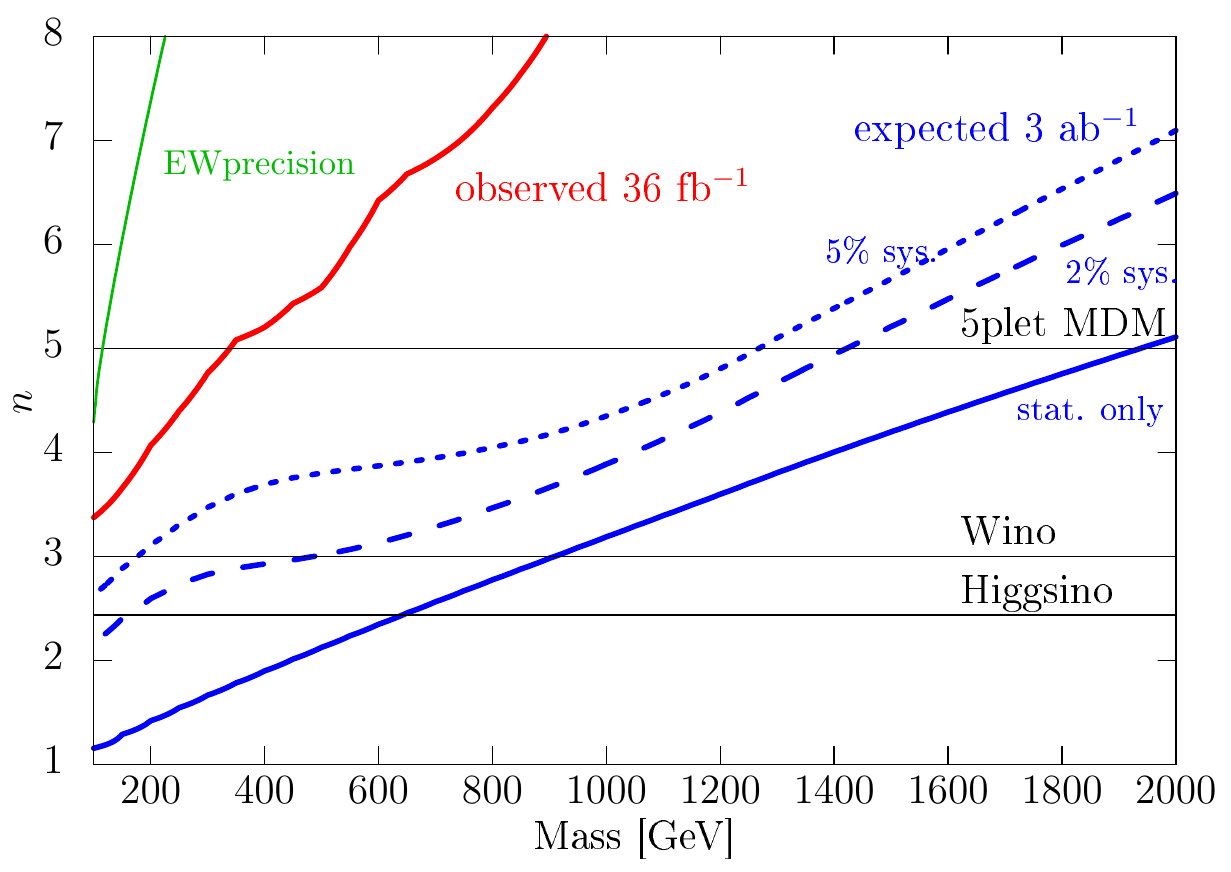}}
	\qquad
	\subcaptionbox{\label{fig:result_scalar} Real scalar}{\includegraphics[width=0.47\textwidth]{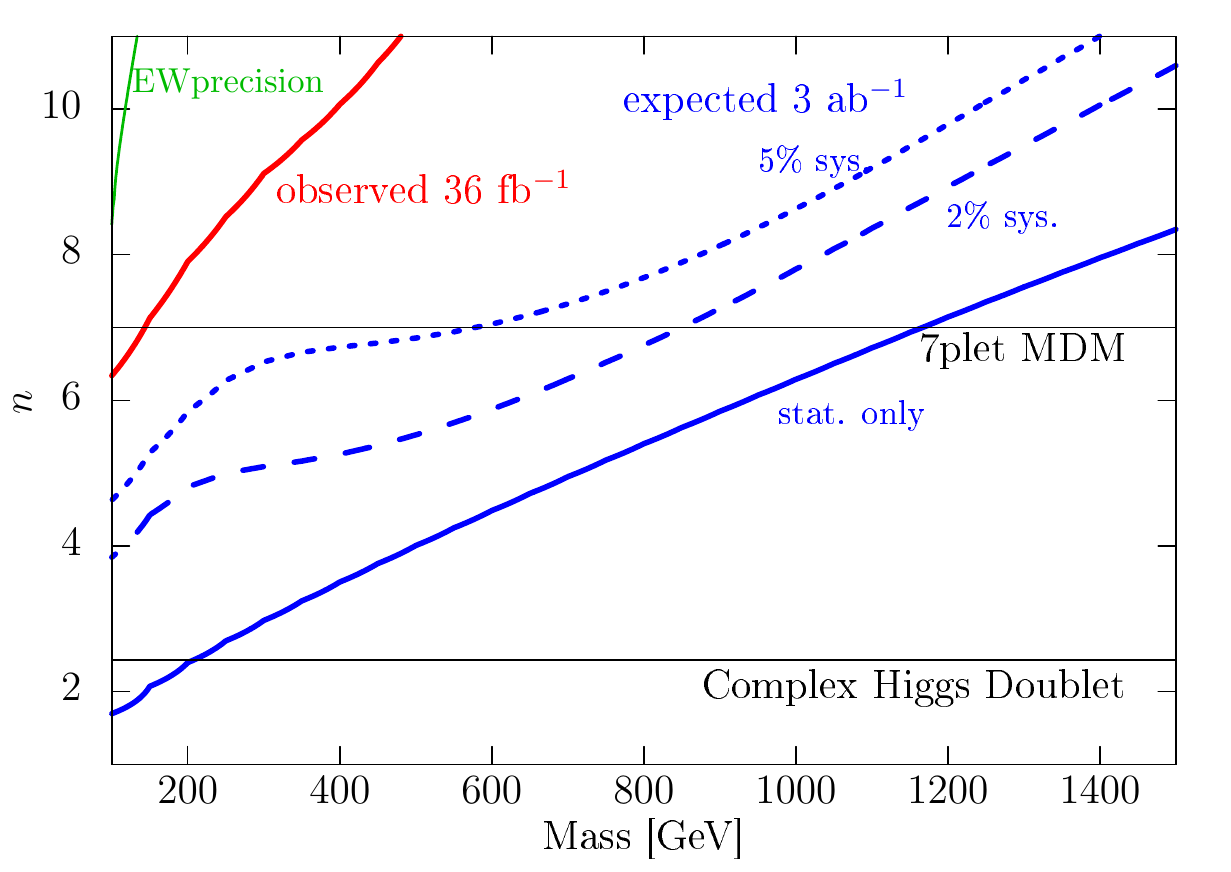}}
\caption{
The current limit and future prospect at 95\% CL for the EWIMP are shown.
The current limit is obtained by using the data at the 13 TeV LHC with the integrated luminosity $36~{\rm fb}^{-1}$ and the future prospect by 14 TeV  and $3~{\rm ab}^{-1}$ LHC.
The region above each line can be experimentally excluded.
The $n$ in the vertical axis shows the effective number of $\SU(2)_{\rm L}$ gauge charge.
}
\label{fig:result}
\end{figure}

In Fig.~\ref{fig:result}, we show the current constraint for Majorana fermion and real scalar from observed data \cite{Aaboud:2017efa} in the red solid lines and future prospect at 14 TeV running LHC with integrated luminosity 3 ${\rm ab}^{-1}$ in the blue lines.
The regions above the lines will be excluded at 95\% CL.
For the cases of a Dirac fermion and complex scalar with  $\SU(2)_{\rm L}$ quantum number $n$ , we can estimate corresponding $n_{\rm corr}$ by solving $n_{\rm corr.}(n_{\rm corr.}^2-1) = 2 n(n^2-1)$, as the overall factor comes from Eq.~\eqref{eq:gaugefactor}.
For the future prospect, we adopt same binning as the current ATLAS study and assume the systematic uncertainty 5\%, 2\% and 0\% for each bin.
The corresponding sensitivity curves are shown in blue dotted, dashed and solid lines in the figure, respectively.
We also show the current constraint from the electroweak precision measurement at the LEP in the green lines  \cite{Matsumoto:2017vfu}.
In the figure, we show the quantum number of several benchmark particles in the horizontal lines.
Note that the $\SU(2)_{\rm L}$ quantum numbers of the Higgsino and complex Higgs doublet 
are estimated to be $n_{\rm corr} \simeq 2.43$, as they are not a Majorana fermion or real scalar.

\section{Conclusion and Discussion}
\label{sec:conclusion}

In this paper, we study the reach of the EWIMP with the precision measurement of the mono-lepton signatures at the current and future LHC.
It is worthwhile comparing the result with our previous work based on dilepton signatures at hadron colliders \cite{Matsumoto:2017vfu}.
If we adopt 2\% systematic uncertainties, we can test around 1.4 TeV 5plet Majorana fermion with the mono-lepton signatures at the future LHC, whereas 1 TeV with dilepton.
In both cases, the sensitivity strongly depends on the systematic uncertainties and simple comparison is dangerous.
However, it seems the mono-lepton will be more useful to probe the EWIMP.
This would be because the SM cross section of mono-lepton process is much greater the dilepton process and the quantum corrections from the EWIMP is relatively larger compared to the dilepton case, since the only left-handed parton quarks and an ${\rm SU}(2)_{\rm L}$ gauge boson contribute the mono-lepton signal.

We found the mono-lepton signature is a quite useful probe of the EWIMP.
However, in order gain more robust constraints and prospects, we need to pay more attention to the systematic uncertainties.
Moreover, if we consider the EWIMP with larger electroweak gauge charges, we need to estimate higher order correction from the EWIMP.
We will leave these issues for a future work.

\section*{Acknowledgments}
This work is supported by Grant-in-Aid for Scientific Research from the Ministry of Education, Culture, Sports, Science, and Technology (MEXT), Japan, No. 16H02176 (S.M.), 17H02878 (S.M. and S.S.),  18K13535 (S.S.) and  26104009 (S.M.) and by World Premier International Research Center Initiative (WPI), MEXT, Japan.

{\it Note Added:} 
During the completion of this paper, a related study
\cite{DiLuzio:2018jwd} was submitted to arXiv which discusses the mono-lepton signature as an indirect probe of the EWIMPs.


\bibliographystyle{aps}
\bibliography{ref}

\end{document}